\documentclass[12pt]{article}
\usepackage{amsfonts}
\usepackage{graphicx}
\usepackage{amsmath}
\usepackage{float}
\textwidth=6.5in \hoffset=-.75in \voffset=-1in \numberwithin{equation}{section}
\numberwithin{figure}{section}

\newcommand {\nn}{\nonumber}
\newcommand {\be}{\begin{equation}}
\newcommand {\ee}{\end{equation}}
\newcommand {\bea}{\begin{eqnarray}}
\newcommand {\eea}{\end{eqnarray}}
\usepackage{amsfonts}
\usepackage{amsmath}
\usepackage{amssymb}
\usepackage{graphicx}
\usepackage{float}
\usepackage{color}

\begin{document}

\begin{titlepage}
\vspace{1cm}
\begin{center}
{\Large \bf {Minimal Surfaces and Generalized Einstein-Maxwell-dilaton Theory}}\\
\end{center}
\vspace{2cm}
\begin{center}
{M. Butler\footnote{mdb815@mail.usask.ca}, A. M. Ghezelbash\footnote{amg142@mail.usask.ca}}
\\
Department of Physics and Engineering Physics, \\ University of Saskatchewan, \\
Saskatoon, Saskatchewan S7N 5E2, Canada\\
\vspace{1cm}
\vspace{2cm}

\end{center}

\begin{abstract}

We present novel classes of non-stationary solutions to the five-dimensional generalized Einstein-Maxwell-dilaton theory with cosmological constant, in which the Maxwell's filed and the cosmological constant couple to the dilaton field. In the first class of solutions, the two non-zero coupling constants are different while in the second class of solutions, the two coupling constants are equal to each other. We find consistent cosmological solutions with positive, negative or zero cosmological constant, where the cosmological constant depends on the value of one coupling constant in the theory. Moreover, we discuss the physical properties of the five-dimensional solutions and the uniqueness of the solutions in five dimensions by showing the solutions with different coupling constants, can't be uplifted to any Einstein-Maxwell theory in higher dimensions.

\end{abstract}
\end{titlepage}\onecolumn 
\bigskip 

\section{Introduction}

One of the main objectives in gravitational physics is to construct, explore and understand the exact solutions to the Einstein field equations in the background of matter fields in different dimensions. 
The possibility of extending the known solutions in asymptotically flat spacetime to the asymptotically de-Sitter and anti de-Sitter solutions, is an important step in better understanding the recent holographic proposals between the extended theories of gravity and the conformal field theories in different dimensions \cite{hol1}, \cite{hol2}. The explored solutions cover a vast area of solutions with different charges, such as  NUT charges \cite{awad}-\cite{holend}, different matter fields, such as Maxwell field, dilaton field \cite{TNme}, \cite{T1} and axion field \cite{T2} and black hole solutions with different horizon topologies \cite{Ch1}-\cite{Jap}.  Moreover, the references \cite{BL1}-\cite{BL45} include the other solutions to extended theories of gravity  with different type of matter fields in different dimensions.  The class of solutions to Einstein-Maxwell-dilaton theory, in which the dilaton field interacts with the cosmological constant and the Maxwell field, was considered in \cite{Torii}, \cite{Kino}. These solutions are relevant to the generalization of the Freund-Rubin compactification of M-theory \cite{FRC}, \cite{FRC2}.

In this article, inspired by construction and exploring the new exact solutions to the field equations of Einstein-Maxwell theory,  we explore and find exact analytical solutions to the five-dimensional Einstein-Maxwell-dilaton theory with cosmological constant and two dilaton coupling constants, based on four-dimensional minimal surfaces. 
A minimal surface is a subset of $ \mathbb{R}^3$ which its mean curvature identically is zero. An alternative definition states that a minimal surface is the critical point of the area functional \cite{Mee12}.
Quite interestingly, Nutku showed that there is a correspondence between gravitational instantons (with anti-self-dual curvature) and any minimal surfaces in $\mathbb{R}^3$ \cite{Nut96}.  
Inspired with the results of \cite{MeK}, we explicitly construct solutions to the Einstein-Maxwell-dilaon theory in five dimensions, where the dilaton field couples to the Maxwell field as well as the cosmological term with two different coupling constants. The solutions are based on four-dimensional Nutku geometry. We solve and show that the field equations support a non-stationary spacetime with non-trivial solutions for the dilaton and Maxwell fields.  Moreover, we consider the case where the two coupling constants are equal and non-zero and find a new class of exact solutions. We discuss the physical properties of the solutions. Finally, we consider the theory where both the coupling constants are zero and show that there is a class of non-trivial solutions to the field equations.  We also show that our new exact solutions can not simply be found as the result of compactification of a higher (than five) dimensional Einstein or Einstein-Maxwell theory with a cosmological constant. 
However, if we consider a higher than five dimensional gravity with a form potential with a cosmological constant, we can compactify the theory to find the generalized Einstein-Maxwell-dilaton theory with two different cosmological constants. 

The paper is organized as follows. 
In section \ref{sec:aneqb}, we consider the minimal surfaces and the generalized Einstein-Maxwell-dilaton theory, in presence of the cosmological constant with two different dilaton coupling constants. We present the four-dimensional Nutku space and use it as the spatial section of the five-dimensional spacetime solutions to the generalized Einstein-Maxwell-dilaton theory. We consider as well, special ansatze for the Maxwell field and the dilaton field. The five-dimensional spacetime has two metric functions. The first metric function depends only on time coordinate, while the second metric function depends on radial and angular coordinates.  After solving the field equations, we find that the the two different coupling constants satisfy a constraint equation. We also find that the cosmological constant can not be any arbitrary quantity, and depends to one of the coupling constants. The cosmological constant then can take any positive, negative or zero value depending on the coupling constant. We discuss the physical properties of the solutions such as the components of the electric field and the dilaton field.
 
 In section \ref{sec:aeqb}, we consider the two coupling constant to be equal in the Einstein-Maxwell-dilaton theory. We consider another set of anastze for the spacetime metyric, the Maxwell field and the dilaton. Specially, one of the metric function depends on the two spatial coordinates as well as the time coordinate. After a long calculation, we consistently solve all the field equations and find the explicit analytical expressions for the metric and other fields in the theory.  Similar to the results in previous section, we find that the cosmological constant takes specific values, depending on the coupling constant in the theory.

In section \ref{sec:aeq0}, we consider the special case, in which both the coupling constants vanish. In this extremal limit of the coupling constants, we notice that the dilaton sector decouples from the theory, hence we recover the Einstein-Maxwell theory.  We employ the same metric ansatz and the gauge field as in section \ref{sec:aeqb}. We find two different exact analytical solutions for the metric function and discuss the properties of the solutions.

In section \ref{sec:novelty}, we discuss the novelty of the solutions and explicitly show that, we can not uplift the solutions with two different coupling constants, into a higher than five dimensional Einstein-Maxwell theory with a cosmological constant. We also show we can not uplift  the solutions with two different coupling constants, into a six-dimensional gravity with a cosmological constant.  Moreover, we show that when the two coupling constants are equal, we can uplift the solutions to a higher than five dimensional Einstein-Maxwell theory only for some specific values of the coupling constant. The cosmological constant of the higher dimensional theory also takes specific values only. 

We close the paper with the concluding remarks and comments for future works.

\section{Exact Solutions to Einstein-Maxwell-dilaton field equations with two different coupling constants $a$ and $b$}
\label{sec:aneqb}

The minimal surfaces describe the minimal area within an existing rigid boundary, such as the surface extended by a soap film  bounded on a wire frame.  Other examples for minimal surfaces include the plane, the helicoid and the catenoid. The helicoid and the catenoid are locally isometric \cite{seven} and are harmonic conjugates of each other \cite{Mee12}. 
The metrics for helicoid and catenoid are given by
\be
ds_{Nutku}^2=\frac{{d{{r}}}^{2}+ \left( \epsilon{N}^{2}+{r}^{2} \right) {d{{\theta}}}^{2}+
	\left( 1+{\epsilon\frac {{N}^{2}  \sin ^2 \theta  
			}{{r}^{2}}} \right) {d{{y}}}^{2}-{\epsilon\frac {{N}^{2}\sin \left( 2\,
			\theta \right) d{{y}}d{{z}}}{{r}^{2}}}+ \left( 1+{\epsilon\frac {{N}^{2}
			 \cos ^2 \theta}{{r}^{2}}} \right) {d
		{{z}}}^{2}
}{ \sqrt{1+{\epsilon\frac {{N}^{2}}{{r}^{2}}}}}\label{Nutku},
\ee
where for the helicoid, $\epsilon=1$, and for the catenoid, $\epsilon=-1$, and we call $N$ as the Nutku parameter. The metric \eqref{Nutku} is asymptotically Euclidean, where the radial coordinate belongs to the interval $[0,+\infty[$ for the helicoid and to the interval $[N,+\infty[$ for the catenoid. The angular coordinate belongs to the interval $0\leq\theta\leq2\pi$ for both the helicoid and catenoid.  
The Ricci scalar and the Ricci tensor of the metric \eqref{Nutku} are identically zero, while the Kretschmann invariant is given by 
\begin{equation}
{\cal K}=\frac{72N^4}{r^4(r^2+\epsilon N^2)^2}+\frac{24a^8}{r^6(r^2+\epsilon N^2)^3},
\end{equation}
respectively.  We notice that the helicoid ($\epsilon=1$) has a singularity at $r=0$, while the catenoid ($\epsilon=-1$), has another singularity at $r=N$. 

We consider a general action for the five-dimensional Einstein-Maxwell-dilaton theory in presence of cosmological constant, in which both Maxwell's filed  and the cosmological constant couple to the dilaton field. The action is
\begin{equation}
S=\int d^5x \sqrt{-g}\{ R-\frac{4}{3}(\nabla \phi)^2-e^{-4/3a\phi}F^2-e^{4/3b\phi}\Lambda\},
\label{act}
\end{equation}
where we consider the gravitational constant to be equal to $(16\pi)^{-1}$.  We consider two different non-zero coupling constants $a$ and $b$ for the coupling of the Maxwell field, as well as the cosmological constant, respectively, to the dilaton field.  We find the Einstein's field equations ${\cal E}_{\mu\nu}=0$, by varying the action \eqref{act} with respect to $g_{\mu\nu}$, where
\begin{equation}
{\cal E}_{\mu\nu}\equiv R_{\mu\nu}-\frac{2}{3}\Lambda g_{\mu\nu}e^{4/3b\phi}-(F_{\mu}^{\lambda}F_{\nu\lambda}-\frac{1}{6}g_{\mu\nu}F^2)e^{-4/3a\phi}-\frac{4}{3}\nabla_\mu \phi \nabla _\nu \phi.\label{einstein}
\end{equation}
We also find the Maxwell's field equations ${\cal M}_{\mu}=0$, by varying the action (\ref{act}) with respect to the Maxwell field $A_\mu$ where 
\begin{equation}
{\cal M}_\mu\equiv\nabla ^\nu (e^{-4/3a\phi}F_{\mu\nu})\label{maxwell}.
\end{equation}
Varying the action (\ref{act}) with respect to the dilaton field, we find the dilaton field equation ${\cal D}=0$ with
\begin{equation}
{\cal D}\equiv \nabla ^2 \phi + \frac{a}{4}e^{-4a\phi/3}F^2-be^{4/3b\phi}\Lambda\label{dilaton}.
\end{equation}
We consider the following ansatz for the five-dimensional metric 
\begin{equation}
ds_5^{2}=-\frac{1}{H^2(r,\theta)}dt^{2}+R^2(t)H(r,\theta)ds_{Nutku}^2,
\label{dsanoteqb}
\end{equation}
where $ds_{Nutku}^2$ is the Nutku line element \eqref{Nutku}.  
We also consider the following form for the Maxwell's field 
\begin{equation}
{A_t}(t,r,\theta)={\alpha R^\gamma(t)}{H^\delta(r,\theta)}\label{gaugeanoteqb}.
\end{equation}
In equations \eqref{dsanoteqb} and \eqref{gaugeanoteqb}, we consider two unknown metric functions $H(r,\theta)$ and $R(t)$, which depend on two coordinates, $r,\,\theta$ and $t$, respectively. We find the values of two constants  $\gamma$ and $\delta$ later, by solving the appropriate filed equations.
The Maxwell's field \eqref{gaugeanoteqb} yields an electric field with components in $r$ and $\theta$ directions. We also consider an ansatz for the dilaton field in terms of metric functions  $H(r,\theta)$ and $R(t)$, such as
\be
\phi(t,r,\theta)=-\frac{3}{4a}\ln(H^\lambda(r,\theta)R^\kappa(t))\label{dilatonanoteqb},
\ee
where $\lambda$ and $\kappa$ are two other constants that we find them later. We find that the Maxwell's equation ${\cal M}^{t}=0$ leads to the differential equation for the metric function $H(r,\theta)$ as
\begin{eqnarray}
&& \left( \delta+\lambda+1 \right)  \left( {\frac {\partial }{\partial 
\theta}}H \left( r,\theta \right)  \right) ^{2}+ \left( {N}^{2}H
 \left( r,\theta \right) \epsilon+{r}^{2}H \left( r,\theta \right) 
 \right) {\frac {\partial ^{2}}{\partial {r}^{2}}}H \left( r,\theta
 \right) \nn\\
 &+& \left( {N}^{2}\delta\,\epsilon+{N}^{2}\epsilon\,\lambda+
\epsilon\,{N}^{2}+\delta\,{r}^{2}+\lambda\,{r}^{2}+{r}^{2} \right) 
 \left( {\frac {\partial }{\partial r}}H \left( r,\theta \right) 
 \right) ^{2}+H \left( r,\theta \right)  \left( {\frac {\partial }{
\partial r}}H \left( r,\theta \right)  \right) r\nn\\
&+&H \left( r,\theta
 \right) {\frac {\partial ^{2}}{\partial {\theta}^{2}}}H \left( r,
\theta \right)=0.
\label{Mt}
\end{eqnarray}

Moreover, we find that the other Maxwell's equations ${\cal M}^r=0,\, {\cal M}^\theta=0$, are satisfied if
\be
\gamma=-\kappa-2. \label{eq1}
\ee
The Einstein's equations ${\cal G}_{tr}=0,\, {\cal G}_{t\theta}=0$ lead to 
\be
\lambda\kappa+4a^2=0,\label{eq2}
\ee
while ${\cal G}_{r\theta}=0$ leads to
\begin{equation}
\lambda+2\delta+2=0,\,\kappa+2\gamma=0,\,4a^2\alpha^2\delta^2=3\lambda^2+6a^2.\label{eq3}
\end{equation}
We find that coefficients $\gamma,\,\delta,\,\lambda$ and $\kappa$ are given by
\be
\gamma=2,\,\delta=-1-\frac{a^2}{2},\,\lambda=a^2,\,\kappa=-4,\label{numvalues}
\ee
while 
\be
\alpha^2=\frac{3}{a^2+2},\label{numvalues2}
\ee 
as the unique solutions to the equations (\ref{eq1}),(\ref{eq2}) and (\ref{eq3}).
We substitute for $\gamma,\,\delta,\,\lambda$ and $\kappa$ in \eqref{Mt}, from equation \eqref{numvalues} and find the partial differential equation 
\bea
&&\left( 2\,{N}^{2}H \left( r,\theta \right) \epsilon+2\,{r}^{2}H
 \left( r,\theta \right)  \right) {\frac {\partial ^{2}}{\partial {r}^
{2}}}H \left( r,\theta \right) + \left( {N}^{2}{a}^{2}\epsilon+{a}^{2}
{r}^{2} \right)  \left( {\frac {\partial }{\partial r}}H \left( r,
\theta \right)  \right) ^{2}\nn\\
&+&2\,H \left( r,\theta \right)  \left( {
\frac {\partial }{\partial r}}H \left( r,\theta \right)  \right) r+
 \left( {\frac {\partial }{\partial \theta}}H \left( r,\theta \right) 
 \right) ^{2}{a}^{2}+2\,H \left( r,\theta \right) {\frac {\partial ^{2
}}{\partial {\theta}^{2}}}H \left( r,\theta \right)=0.
\eea

To solve this differential equation, we first try to separate the differential equation. Although separating the differential equation leads to two ordinary differential equations, however the solutions are very complicated and not suitable for determining the metric function $H(r,\theta)$.  Analyzing the separated solutions leads to considering a change of function from 
 $H(r,\theta)$ to a new function $K(r,\theta)$, which is given by 
\be
K(r,\theta)=H(r,\theta)^{a^2+{2}}.\label{CF}
\ee
Using the change of function \eqref{CF}, we find the following partial differential equation for $K(r,\theta)$, as 
\bea
&&\left( K \left( r,\theta \right)  \right) ^{-{\frac {2\,{a}^{2}+3}{{a
}^{2}+2}}}({N}^{2}\epsilon+r^2)\, \left( {\frac {\partial }{\partial r}}K
 \left( r,\theta \right)  \right) ^{2}-2\, \left( K \left( r,\theta
 \right)  \right) ^{-{\frac {{a}^{2}+1}{{a}^{2}+2}}}({N}^{2}\epsilon+r^2)\,{
\frac {\partial ^{2}}{\partial {r}^{2}}}K \left( r,\theta \right) \nn\\
&+& \left( K \left( r,
\theta \right)  \right) ^{-{\frac {2\,{a}^{2}+3}{{a}^{2}+2}}} \left( {
\frac {\partial }{\partial \theta}}K \left( r,\theta \right)  \right) 
^{2}-2\, \left( K \left( r,\theta \right)  \right) ^{-{\frac {{a}^{2}+
1}{{a}^{2}+2}}} \left( {\frac {\partial }{\partial r}}K \left( r,
\theta \right)  \right) r-2\, \left( K \left( r,\theta \right) 
 \right) ^{-{\frac {{a}^{2}+1}{{a}^{2}+2}}}{\frac {\partial ^{2}}{
\partial {\theta}^{2}}}K \left( r,\theta \right)=0.\nn\\
&&
\label{Keq}
\eea

Quite interestingly, we find exact solutions to \eqref{Keq} as
\be
K(r,\theta)=(1+\frac{k_1 \, r \, \sin \theta}{N}+\frac{k_2 \, r \, \cos \theta}{N})^2,\label{GG}
\ee
where $k_1$ and $k_2$ are two constants. 

\begin{figure}[H]
\centering
\includegraphics[width=0.4\textwidth]{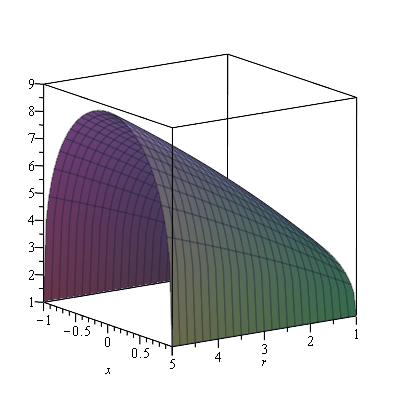}
\caption{The metric function $H(r,\theta)$ as function of $r$ and $x=\cos\theta$, where we set $k_1=2,\,k_2=3,\,N=1,\,a=1$.}
\label{fig1}
\end{figure}

To show the typical behavior of the metric function $H(r,\theta)$, we plot it as a function of $r$ and $x=\cos\theta$ in figure \ref{fig1}, where we set $k_1=2,\,k_2=3,\,N=1,\,a=1$.

We note that substituting for $\alpha$ from equation \eqref{numvalues2} along with $H(r,\theta)$ from \eqref{CF} and \eqref{GG} leads to ${\cal G}_{tr}=0$, ${\cal G}_{t\theta}=0$ and ${\cal G}_{r\theta}=0$.  We then consider the other Einstein's field equation ${\cal G}_{tt}=0$ and solve it to find the cosmological constant  $\Lambda$ in terms of the time-dependent metric function $R(t)$.  We then substitute the resultant equation for the cosmological constant in Einstein's field equation equation ${\cal G}_{rr}=0$. We finally get an ordinary differential equation for $R(t)$ such as
\be
a^2R(t)\frac{d^2R(t)}{dt^2}=(a^2-4)(\frac{dR(t)}{dt})^2.\label{eqforR}
\ee
We find the solutions to the differentia equation (\ref{eqforR}) are given by
\be
R(t)=(\eta\,t+\zeta)^{a^2/4},\label{RRR}
\ee
where $\eta$ and $\zeta$ are two constants.
We substitute the time-dependent function f$R(t)$ as given by (\ref{RRR}) into the equation for the cosmological constant $\Lambda$ where we found earlier. This gives an equation for the cosmological constant as
\be
\Lambda=\frac{3a^2(a^2-1)\eta^2}{8}\frac{K(r,\theta)^{\frac{ab+2}{a^2+2}}}{(\eta\,t+\zeta)^{ab+2}},\label{LamQ}
\ee
where $K(r,\theta)$ is given in (\ref{GG}).

Of course, to get a coordinate-independent cosmological constant from the right hand side of \eqref{LamQ}, we should set
\be
ab=-2\label{ab},
\ee
where we find
\be
\Lambda=\frac{3a^2(a^2-1)\eta^2}{8}.\label{cosmo}
\ee
Depending on coupling constant, the cosmological constant in the Einstein-Maxwell-dilaton theory can be positive, negative and even zero, as we notice from equation (\ref{cosmo}). We also notice that the constraint on the coupling constants (\ref{ab}) as well as the cosmological constant (\ref{LamQ}) are independent of choice of $\epsilon$. 

We substitute the constraint (\ref{ab}) as well as the cosmological constant \eqref{cosmo}, in all the other remaining Einstein's field equations and find that all equations are satisfied. Moreover, we substitute all the known variables and the metric functions in the dilaton field equation \eqref{dilaton} and find that it is exactly satisfied. 

We conclude that the five-dimensional ansatz for the spacetime (\ref{dsanoteqb}) becomes
\begin{eqnarray}
ds_5^2&=& -(1+\frac{k_1 \, r \, \sin \theta}{N}+\frac{k_2 \, r \, \cos \theta}{N})^{-\frac{4}{a^2+2}}dt^2\nn\\
&+&\frac{(\eta t+\zeta)^{\frac{a^2}{2}}}{\sqrt{1+{\epsilon\frac {{N}^{2}}{{r}^{2}}}}}
(1+\frac{k_1 \, r \, \sin \theta}{N}+\frac{k_2 \, r \, \cos \theta}{N})^{\frac{2}{a^2+2}}\nn\\
&\times&\big(
{d{{r}}}^{2}+ \left( \epsilon{a}^{2}+{r}^{2} \right) {d{{\theta}}}^{2}+
	\left( 1+{\epsilon\frac {{N}^{2}  \sin^2\theta 
			}{{r}^{2}}} \right) {d{{y}}}^{2}-{\epsilon\frac {{N}^{2}\sin \left( 2\,
			\theta \right) d{{y}}d{{z}}}{{r}^{2}}}+ \left( 1+{\epsilon\frac {{N}^{2}
			\cos^2\theta}{{r}^{2}}} \right) {d
		{{z}}}^{2}
\big).\nn\\
&&
\label{metric5danoteqb}
\end{eqnarray}
The Ricci scalar as well as the Kretschmann invariant of the spacetime (\ref{metric5danoteqb}) are both divergent at $r=0$ for $\epsilon=1$ and at $r=0,N$ for $\epsilon=-1$, on the singular points of the Nutku geometry (\ref{Nutku}) and also on the hypersurface $N+k_1r\sin\theta+k_2r\cos\theta=0$. In fact, the same behaviours were noticed in higher-dimensional supergravity solutions, based on transverse self-dual hyper-K\"ahler manifolds \cite{36}-\cite{38}. We expect that all the irregular hypersurfaces of our solutions can be avoided and changed to regular horizon hypersurfaces, if we consider more coordinate dependence in the metric functions \cite{39}. Moreover for $r\rightarrow \infty$, the Ricci and Kretschmann scalars are divergent, however if we rescale the metric (\ref{metric5danoteqb}), according to 
\be
d\tilde s_5^2=e^{-\frac{4}{3}a\phi}ds_5^2,
\ee
we find that the metric is regular asymptotically, for the dilaton coupling constant $a^2 \geq 2$.  Moreover, we choose $\eta \zeta >0$ to avoid any singularity on time slice $t=-\frac{\zeta}{\eta}$, where the time coordinate $t$ runs from $0$ to $\infty$.

\begin{figure}[H]
\centering
\includegraphics[width=0.4\textwidth]{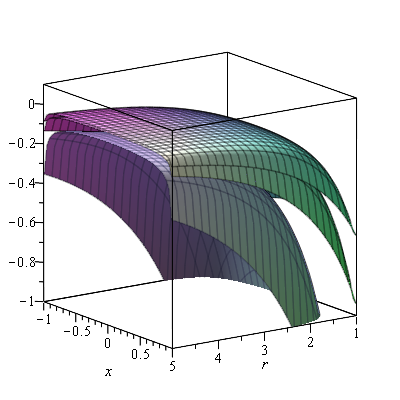}

\caption{The $r$-component of electric field as function of $r$ and $x=\cos\theta$ for three time slices; $t=1$ (upper surface), $t=10$ (middle surface) and $t=100$ (lower surface), where we set $k_1=2,\,k_2=3,\,N=1,\,a=1,\,\eta=1,\,\zeta=5$.}
\label{fig2}
\end{figure}

\begin{figure}[H]
\centering
\includegraphics[width=0.4\textwidth]{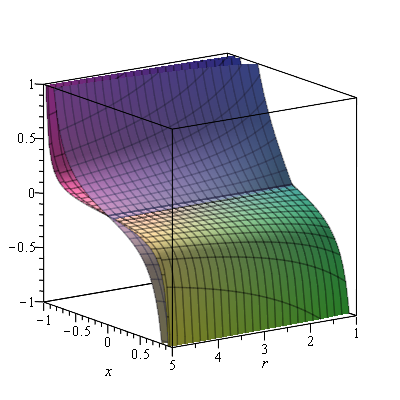}

\caption{The $\theta$-component of electric field 
as function of $r$ and $x=\cos\theta$ for three time slices; $t=1$ (lower surface), $t=10$ (middle surface) and $t=100$ (upper surface), where we set $k_1=2,\,k_2=3,\,N=1,\,a=1,\,\eta=1,\,\zeta=5$.}
\label{fig3}
\end{figure}

The Maxwell's field \eqref{gaugeanoteqb}, explicitly read as
\be
A_t(t,r,\theta)=\sqrt{\frac{3}{a^2+2}}\,\frac{N(\eta\,t+\zeta)^{a^2/2}}{N+k_1 \, r \, \sin \theta+k_2 \, r \, \cos \theta},\label{Atexp}
\ee
that leads to an electric field in $r$ and $\theta$ directions. 
We present the electric field in $r$ and $\theta$ directions for different time slices as functions of $r$ and $\theta$, in figures \ref{fig2} and \ref{fig3}.

\begin{figure}[H]
\centering
\includegraphics[width=0.4\textwidth]{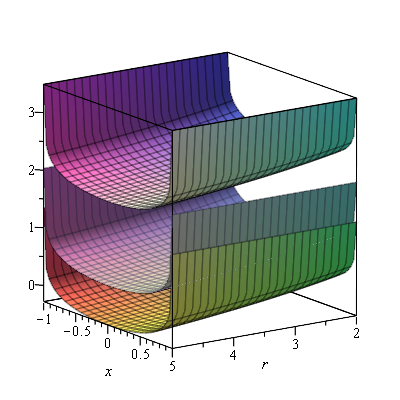}
\caption{The dilaton field $\phi(t,r,\theta)$ as function of $r$ and $x=\cos\theta$ for three time slices; $t=1$ (lower surface), $t=10$ (middle surface) and $t=100$ (upper surface), where we set $k_1=2,\,k_2=3,\,N=1,\,a=1,\,\eta=1,\,\zeta=5$.}
\label{fig4}
\end{figure}

We also find that the dilaton field \eqref{dilatonanoteqb}, explicitly is given by
\be
\phi(t,r,\theta)={\frac{-3a}{4}}\,\ln \frac{(1+\frac{k_1 \, r \, \sin \theta}{N}+\frac{k_2 \, r \, \cos \theta}{N})^{\frac{2}{2+a^2}}}{(\eta\,t+\zeta)}. \label{phiexp}
\ee
We present in figure \ref{fig4} the typical behaviour of the dilaton field, for different time slices, as a function of $r$ and $\theta$, where we set $k_1=2,\,k_2=3,\,N=1,\,a=1,\,\eta=1,\,\zeta=5$. 

\section{Exact Solutions with equal coupling constants $a$ and $b$}
\label{sec:aeqb}
In this section, we consider the Einstein-Maxwell-dilaton theory, where the coupling constants $a$ and $b$ are equal to each other. The action is given by
\begin{equation}
S=\int d^5x \sqrt{-g}\{ R-\frac{4}{3}(\nabla \phi)^2-e^{-4/3a\phi}(F^2+\Lambda)\}.
\label{act2}
\end{equation}
We consider the five-dimensional metric as
\begin{equation}
ds_5^{2}=-\frac{1}{H^2(t,r,\theta)}dt^{2}+R^2(t)H(t,r,\theta)ds_{Nutku}^2,
\label{dsaeqb}
\end{equation}
where $ds_{Nutku}^2$ is given by (\ref{Nutku}). The main difference between the metric ansatze  (\ref{dsanoteqb}) and (\ref{dsaeqb}) is that, in the former case, there is no time dependence for the metric function $H$, while in the latter case, the metric function $H$ depends explicitly on the time coordinate as well as two spatial coordinates.  We also use the following ansatz for the dependence of the electromagnetic field on the coordinates 
\be
{A_t}(t,r,\theta)={\alpha}{ R^\gamma(t)H^\delta(t,r,\theta)}\label{gaugeaeqb},
\ee
where $\gamma$ and $\delta$ are two constants that we will find them later.
Also, we consider a dilaton field that depends on both metric functions $H(t,r,\theta)$ and $R(t)$, and is given by
\be
\phi(t,r,\theta)=-\frac{3}{4a}\ln(R^\lambda(t)H^\kappa(t,r,\theta)),\label{phicase2}
\ee
where $\lambda$ and $\kappa$ are two other constants.  Due to extra terms arising from the time derivative of metric function $H(t,r,\theta)$, finding solutions to the field equations are more tedious than chapter \ref{sec:aneqb}. We find that 
the three non-zero Maxwell's equations are given by
\bea
{{\cal M}^t}{ }&=&H \left( t,r,\theta \right)  \left( {N}^{2}\epsilon+{r}^{2} \right) {
\frac {\partial ^{2}}{\partial {r}^{2}}}H \left( t,r,\theta \right) +H
 \left( t,r,\theta \right) {\frac {\partial ^{2}}{\partial {\theta}^{2
}}}H \left( t,r,\theta \right) \nn\\
&+& \left( {N}^{2}\epsilon+{r}^{2}
 \right)  \left( \delta+\kappa+1 \right)  \left( {\frac {\partial }{
\partial r}}H \left( t,r,\theta \right)  \right) ^{2}+ \left( {\frac {
\partial }{\partial r}}H \left( t,r,\theta \right)  \right) H \left( t
,r,\theta \right) r\nn\\
&+& \left( {\frac {\partial }{\partial \theta}}H
 \left( t,r,\theta \right)  \right) ^{2} \left( \delta+\kappa+1
 \right)=0,
\label{EQ1}
\eea
\bea
{\cal M}^r&=&\left( {\frac {\partial ^{2}}{\partial t\partial r}}H \left( t,r,
\theta \right)  \right) R \left( t \right) H \left( t,r,\theta
 \right) \nn\\
 &+& \left( R \left( t \right)  \left( \delta+\kappa+1 \right) {
\frac {\partial }{\partial t}}H \left( t,r,\theta \right) + \left( {
\frac {\rm d}{{\rm d}t}}R \left( t \right)  \right) H \left( t,r,
\theta \right)  \left( {\gamma}+\lambda+2 \right)  \right) {\frac 
{\partial }{\partial r}}H \left( t,r,\theta \right)=0,\nn\\
&&
\label{EQ2}
\eea
\bea
{\cal M}^\theta&=&\left( {\frac {\partial ^{2}}{\partial \theta\partial t}}H \left( t,r
,\theta \right)  \right) R \left( t \right) H \left( t,r,\theta
 \right) \nn\\
 &+& \left( R \left( t \right)  \left( \delta+\kappa+1 \right) {
\frac {\partial }{\partial t}}H \left( t,r,\theta \right) + \left( {
\frac {\rm d}{{\rm d}t}}R \left( t \right)  \right) H \left( t,r,
\theta \right)  \left( {\gamma}+\lambda+2 \right)  \right) {\frac 
{\partial }{\partial \theta}}H \left( t,r,\theta \right)=0.\nn\\
&&
\label{EQ3}
\eea

\begin{figure}[H]
\centering
\includegraphics[width=0.4\textwidth]{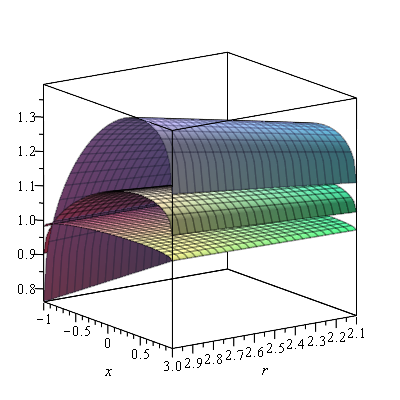}

\caption{The metric function $H(t,r,\theta)$ as function of $r$ and $x=\cos\theta$  for three time slices $t=1$ (upper surface), $t=2$ (middle surface) and $t=5$ (lower surface), where we set $N=2,\,l_1=5,\,l_2=3,\,a=1,\tau=1,\xi=2$.}
\label{fig5}
\end{figure}

\begin{figure}[H]
\centering
\includegraphics[width=0.4\textwidth]{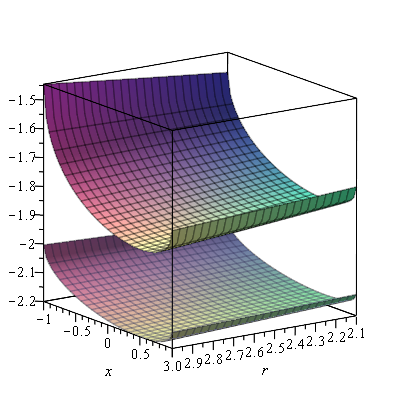}

\caption{The dilaton $\phi(t,r,\theta)$ as function of $r$ and $x=\cos\theta$  for two time slices $t=1$ (upper surface) and $t=2$ (lower surface), where we set $N=2,\,l_1=5,\,l_2=3,\,a=1,\tau=1,\xi=2$.}
\label{fig6}
\end{figure}

\begin{figure}[H]
\centering
\includegraphics[width=0.4\textwidth]{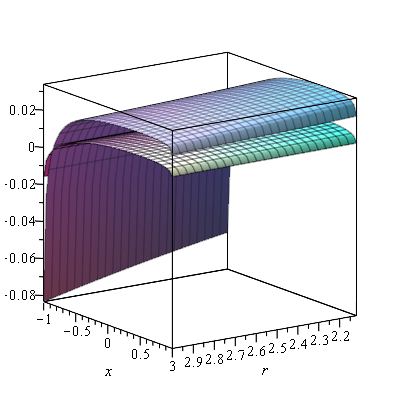}

\caption{The $r$-component of electric field 
as function of $r$ and $x=\cos\theta$ for two time slices; $t=1$ (upper surface), $t=2$ (lower surface), where we set $N=2,\,l_1=5,\,l_2=3,\,a=1,\tau=1,\xi=2$.}
\label{fig7}
\end{figure}

\begin{figure}[H]
\centering
\includegraphics[width=0.4\textwidth]{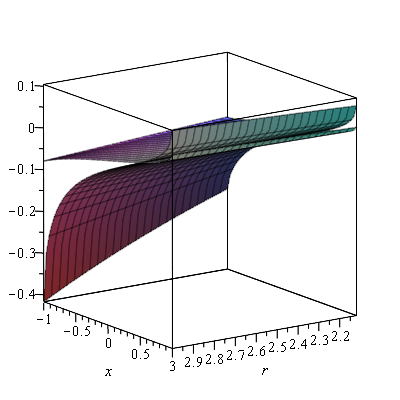}

\caption{The $\theta$-component of electric field 
as function of $r$ and $x=\cos\theta$ for twp time slices; $t=1$ (lower surface), $t=2$ (upper surface), where we set $N=2,\,l_1=5,\,l_2=3,\,a=1,\tau=1,\xi=2$.}
\label{fig8}
\end{figure}

We find $\kappa$ from the equation (\ref{EQ1}) and substitute for it in the other two equations (\ref{EQ2}) and (\ref{EQ3}).  We find $\lambda$ from solving each of these two equations. The two solutions for $\lambda$ are equal to each other if the following equation satisfies
\be
\frac{\partial ^2 H}{\partial t\partial r}\frac{\partial H}{\partial \theta}=\frac{\partial ^2 H}{\partial t\partial \theta}\frac{\partial  H}{\partial r}.\label{cond}
\ee 
The above equation (\ref{cond}) yields that $\frac{\partial H}{\partial r}-\frac{\partial H}{\partial \theta}$ must be only a function of coordinates $r$ and $\theta$.  We then choose the metric function $H(t,r,\theta)$ as
\be
H(t,r,\theta)=R^{-2}(t)\{R^\nu(t)+L(r,\theta)\}^\omega,\label{Hansatz}
\ee
to satisfy $\frac{\partial H}{\partial r}-\frac{\partial H}{\partial \theta}$ being a function of $r$ and $\theta$. In \eqref{Hansatz},  $\omega$ and $\nu$ are two constants and  $L$ is an arbitrary function of $r$ and $\theta$.
The  Einstein's equation ${\cal G}^{tr}=0$ and  ${\cal G}^{t\theta}=0$ lead to following constraints on constants after substituting equation (\ref{Hansatz}) for $H$ 
\be
\lambda=2\kappa,\,\omega (\kappa^2+2a^2)=2 a^2.\label{UW}
\ee
We then substitute equations (\ref{Hansatz}) and equation (\ref{UW}), in the Maxwell's equations ${\cal M}^{r}=0,\,{\cal M}^\theta=0$, and find two other constraints on the constants, such as 
\be
\nu(-2\kappa a^2-2\delta a^2+\kappa^2-2a^2)+(\gamma-2\delta)(-\kappa^2-2a^2)+2\kappa^2+4a^2=0,\label{Mr1}
\ee
and
\be
(\kappa^2+2a^2)(\gamma-2\delta-2)=0.\label{Mr2}
\ee
We notice that we can't consider $\kappa^2=-2a^2$ in (\ref{Mr2}), because the second equation of  (\ref{UW}), then enforces a divergent value for $\omega$. We find from equation (\ref{Mr2}) that 
\be 
\delta=\frac{\gamma}{2}-1.
\ee
Moreover, the Einstein's equation ${\cal G}^{r\theta}=0$, implies two constraints as 
\be
\kappa+2\delta+2=0\label{Grth1},
\ee
and 
\be
4\alpha^2\delta^2a^2=3\kappa^2+6a^2\label{Grth2}.
\ee
The equation \eqref{Grth1} implies $\gamma=-\kappa$ and equation \eqref{Grth2} gives $\alpha^2$, in terms of $\kappa$ as
\be
\alpha^2=\frac{3\kappa^2+6a^2}{(2+\kappa)^2a^2}.
\ee
Simplifying equation ${\cal G}^{r\theta}=0$ leads to finding $\kappa$ as
\be
\kappa=a^2.
\ee
Moreover, from equation (\ref{UW}), we get 
\bea
\lambda&=&2a^2,\,\\
\omega&=&\frac{2}{2+a^2}.
\eea
From equation (\ref{Grth1}), we also find that $\delta$ and $\gamma$ are related to the coupling constant $a$, by
\be
\delta=-1-\frac{a^2}{2},
\ee
and 
\be
\gamma=-a^2,
\ee
respectively. We should notice that equation \eqref{Mr1} does not lead to any definite value for the constant $\nu$.
So, we summarize the results, so far, for the different fields in terms of only two metric functions $R(t)$ and $L(r,\theta)$ as well as the constant $\nu$. The Maxwell's field and the dilaton field are given by
\be
A_t(t,r,\theta)=\frac{3}{a^2+2}R^{-a^2}(t)H^{-1-a^2/2}(t,r,\theta),\label{Afix}
\ee
and
\be
\phi(t,r,\theta)=\frac{-3}{4a}\ln(R^{2a^2}(t)H^{a^2}(t,r,\theta)),\label{dilfix}
\ee
respectively, where 
\be
H(t,r,\theta)=R^{-2}(t)\{R^{ {\nu}}(t)+L(r,\theta)\}^{\frac{2}{2+a^2}}.\label{HGfix}
\ee
To find the metric function $R(t)$ and the constant $\nu$, we solve the Einstein's equation ${\cal G}_{tt}=0$ for the cosmological constant and then
substitute the result in the other Einstein's equation ${\cal G}_{rr}=0$. We find the following differential equation for the metric function $R(t)$ 
\be
\left( R \left( t \right)  \right) ^{4\,{a}^{2}+2} \left( {\frac 
{\rm d}{{\rm d}t}}R \left( t \right)  \right) ^{2}({a}^{2}-1)
+ \left( R \left( t \right) 
 \right) ^{4\,{a}^{2}+3}{\frac {{\rm d}^{2}}{{\rm d}{t}^{2}}}R \left( 
t \right)=0,
\label{E11}
\ee
We find that the solutions for $R(t)$ to the differential equation (\ref{E11}) are given by
\be
R(t)=(\tau t+\xi)^{\frac{1}{a^2}},\label{RR}
\ee
where $\tau$ and $\xi$ are two constants.
We also find that function $L(r,\theta)$ shall satisfy the partial differential equations
\bea 
&& L \left( r,\theta \right)  \left( {N}^{2}\epsilon+{r}^{2} \right) {
\frac {\partial ^{2}}{\partial {r}^{2}}}L \left( r,\theta \right) +
 \left( {\frac {\partial ^{2}}{\partial {\theta}^{2}}}L \left( r,
\theta \right)  \right) L \left( r,\theta \right) \nn\\
&+& \left( -1+ \left( 
1/3\,{a}^{2}+2/3 \right) {\alpha}^{2} \right)  \left( {\frac {
\partial }{\partial \theta}}L \left( r,\theta \right)  \right) ^{2}+rL
 \left( r,\theta \right) {\frac {\partial }{\partial r}}L \left( r,
\theta \right)=0,
\label{E21}
\eea
and
\be
\left( {N}^{2}\epsilon+{r}^{2} \right) {\frac {\partial ^{2}}{
\partial {r}^{2}}}L \left( r,\theta \right) +r{\frac {\partial }{
\partial r}}L \left( r,\theta \right) +{\frac {\partial ^{2}}{
\partial {\theta}^{2}}}L \left( r,\theta \right)=0,
\label{E22}
\ee
contingent to choosing 
\be
\nu=2+a^2.
\ee
Quite interestingly, we find the solutions to the differential equations  (\ref{E21}) and (\ref{E22})  as 
\be
L(r,\theta)=1+{l_1}\frac{r\cos\theta}{N}+{l_2}\frac{r\sin\theta}{N},\label{GG1}
\ee
where $l_1$ and $l_2$ are two constants.
We notice that the function $L(r,\theta)$ resembles $K(r,\theta)$ in equation (\ref{GG}). After substituting for the known functions $R(t)$, $L(r,\theta)$ and $H(t,r,\theta)$ in the equation for the cosmological constant, we  find that
\be
\Lambda=\frac{3\tau^2}{2}\frac{4-a^2}{a^4}.\label{CosmoC}
\ee
The cosmological constant $\Lambda$ can be positive, zero or negative, depending on the coupling constant $a<2$, $a=2$ or $a>2$, respectively.  We also verify that all other remaining field equations, including the dilaton field equation, are satisfied by substituting the known metric functions and the cosmological constant. We plot the typical behaviour of $H(t,r,\theta)$ and the dilaton field $\phi(t,r,\theta)$  in figures \ref{fig5} and \ref{fig6} on some $t={\text {constant}}$ hyper-surfaces. We also plot the components of the electric field in figures \ref{fig7} and \ref{fig8} on the same $t={\text {constant}}$ hyper-surfaces. In all plots, we choose the specific values for the constants $N=2,\,l_1=5,\,l_2=3,\,a=1,\tau=1,\xi=2$.

We show in section \ref{sec:novelty} that the five-dimensional metric 
\begin{eqnarray}
ds_5^2&=& -(\tau t+\xi)^{\frac{4}{a^2}}\{(\tau t+\xi)^{\frac{2+a^2}{a^2}}+{l_1}\frac{r\cos\theta}{N}+{l_2}\frac{r\sin\theta}{N}\}^{-4/(a^2+2)}dt^2\nn\\
&+&\frac{\{(\tau t+\xi)^{\frac{2+a^2}{a^2}}+{l_1}\frac{r\cos\theta}{N}+{l_2}\frac{r\sin\theta}{N}\}^{2/(a^2+2)}}{\sqrt{1+{\epsilon\frac {{N}^{2}}{{r}^{2}}}}}       
\nn\\
&\times&\big(
{{d{{r}}}^{2}+ \left( \epsilon{N}^{2}+{r}^{2} \right) {d{{\theta}}}^{2}+
	\left( 1+{\epsilon\frac {{N}^{2} \sin^2 \theta 
			{}}{{r}^{2}}} \right) {d{{y}}}^{2}-{\epsilon\frac {{N}^{2}\sin \left( 2\,
			\theta \right) d{{y}}d{{z}}}{{r}^{2}}}+ \left( 1+{\epsilon\frac {{N}^{2}
			 \cos ^2\theta }{{r}^{2}}} \right) {d
		{{z}}}^{2}
}
\big),\nn\\
&&
\label{metraeqb}
\end{eqnarray}
can be uplifted to the solutions of the six or higher dimensional  Einstein-Maxwell theory with a cosmological constant, for specific values of the coupling constant $a$.


\section{Solutions with both coupling constants $a=b$ equal to zero}
\label{sec:aeq0}

\begin{figure}[H]
\centering
\includegraphics[width=0.4\textwidth]{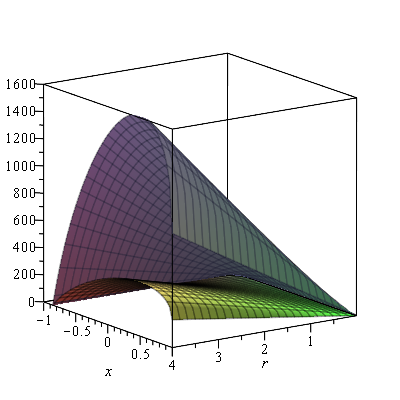}
\caption{The metric function $H_+(t,r,\theta)$  
as a function of $r$ and $x=\cos\theta$ for two time slices $t=2$ (lower surface) and $t=3$ (upper surface), where we set $N=2,\,f_1=5,\,f_2=3,\,\Lambda=3$.}
\label{fig9}
\end{figure}

\begin{figure}[H]
\centering
\includegraphics[width=0.4\textwidth]{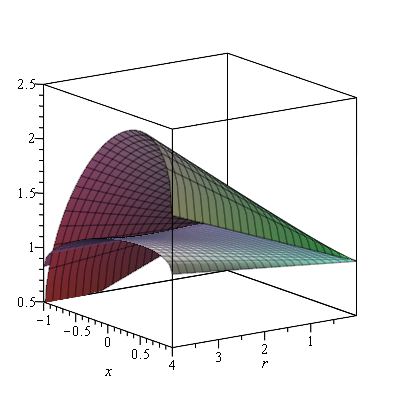}
\caption{The metric function $H_-(t,r,\theta)$  
as a function of $r$ and $x=\cos\theta$ for two time slices $t=2$ (upper surface) and $t=3$ (lower surface), where we set $N=2,\,f_1=5,\,f_2=3,\,\Lambda=3$.}
\label{fig10}
\end{figure}

In this section, we consider a special case, where both coupling constants are zero. We note that we simply can't consider the solutions of the previous section and substitute the coupling constant $a=b \rightarrow 0$. This leads to the diverging dilaton field (\ref{dilfix}) and the cosmological constant (\ref{CosmoC}). In the limit of $a=b\rightarrow 0$, we notice that the dilaton field decouples from the theory (with the action (\ref{act})) and the theory simplifies to the Einstein-Maxwell theory withca cosmological constant.  We consider the metric ansatz 
\begin{equation}
ds_5^{2}=-\frac{1}{H(t,r,\theta)^{2}}dt^{2}+R(t)^2H(t,r,\theta)ds_{EH}^2,
\label{dsaeqbzero}
\end{equation}
exactly the same as (\ref{dsaeqb}). We also consider an ansatz for the Maxwell field, such as
\be
{A_t}(t,r,\theta)=\frac{\alpha}{H(t,r,\theta)}\label{gaugeaeqbzero}.
\ee
Inspired with the solutions (\ref{GG}) and (\ref{GG1}) for the metric function $H$, where the coupling constants are not zero, we consider a possible solution for the metric function $H(t,r,\theta)$ as
\be
H(t,r,\theta)=1+\frac{{f_1}{r\cos\theta}+{f_2}{r\sin\theta}}{R(t)^2},\label{Hansatzabzero}
\ee
where $f_1$ and $f_2$ are two constants. The Maxwell's equations and some non-diagonal Einstein's equations lead to
\be
\alpha^2=\frac{3}{2}.
\ee 
To find the time-dependent metric function $R(t)$, we symbolically solve the equation ${\cal G}_{\phi\psi}=0$ for the cosmological constant $\Lambda$. We substitute the result for the cosmplogical constant in  ${\cal G}_{tt}=0$ and find 
\be
R(t)\frac{d^2 R(t)}{dt^2}-(\frac{d R(t)}{dt})^2=0.\label{Reqabzero}
\ee
The solutions to \eqref{Reqabzero} are
\be
R(t)=R_0e^{\psi t},\label{Rsolabzero}
\ee
where $R_0$ and $\psi$ are two constants.  Moreover, from equation ${\cal G}_{\phi\psi}=0$, we find that
\be
\psi=\pm \sqrt{\frac{\Lambda}{6}}.
\ee
So, we find the metric function $H(t,r,\theta)$ is given by
\be
H_{\pm}(t,r,\theta)=1+e^{\pm 2 \sqrt{\frac{\Lambda}{6}}t}({f_1}{r\cos\theta}+{f_2}{r\sin\theta}).\label{Hansatzabzerofinal}
\ee
We explicitly check that all the other remaining Maxwell's and Einstein's equations are satisfied after substituting the above results. 
 Figures \ref{fig9} and \ref{fig10} show the behaviour of the metric function $H_+$ and $H_-$ , respectively,  as a function of $r$ and $x=\cos\theta$, where we set $N=2,\,f_1=5,\,f_2=3,\,\Lambda=3$.
We note that the metric function $H_+$ monotonically increases by increasing the time coordinate. On the other hand, the metric function $H_-$ monotonically decreases by increasing the time coordinate. 
 
\begin{figure}[H]
\centering
\includegraphics[width=0.4\textwidth]{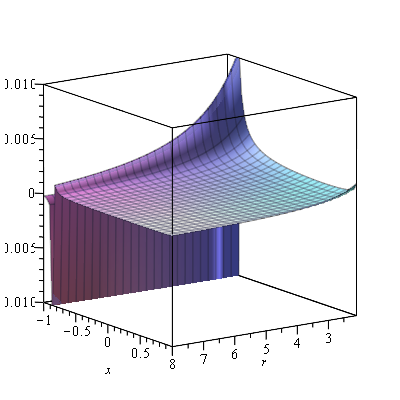}
\includegraphics[width=0.4\textwidth]{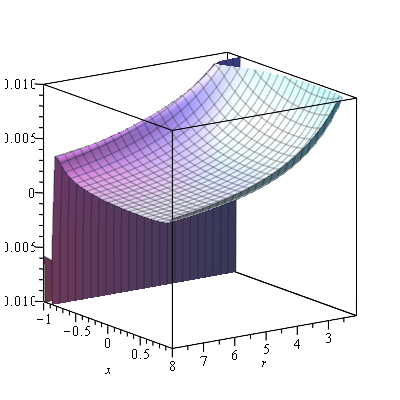}
\caption{The $r$-component of electric field 
as function of $r$ and $x=\cos\theta$ for $t=1$, with $H_+$ (left) and $H_-$ (right), where we set $N=2,\,f_1=5,\,f_2=3,\,\Lambda=3$.}
\label{fig11}
\end{figure}

\begin{figure}[H]
\centering
\includegraphics[width=0.4\textwidth]{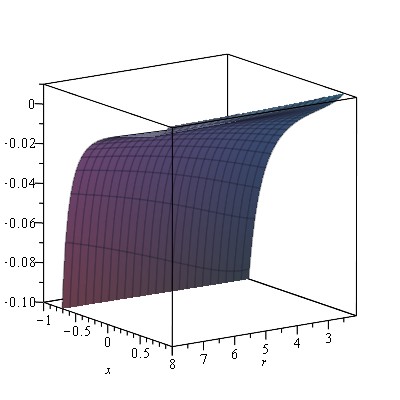}
\includegraphics[width=0.4\textwidth]{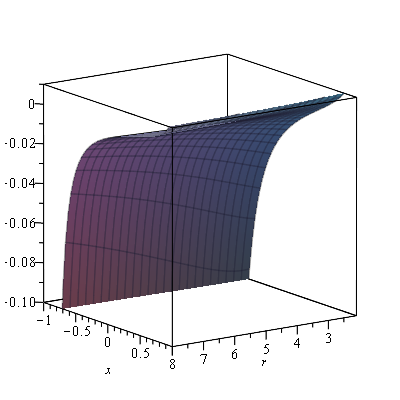}
\caption{The $\theta$-component of electric field 
as function of $r$ and $x=\cos\theta$ for $t=1$, with $H_+$ (left) and $H_-$ (right), where we set $N=2,\,f_1=5,\,f_2=3,\,\Lambda=3$.}
\label{fig12}
\end{figure}

In figures \ref{fig11} and \ref{fig12}, we plot the components of the electric field in terms of coordinates $r$ and $x=\cos\theta$,  where we set $N=2,\,f_1=5,\,f_2=3,\,\Lambda=3$.
 
\section{Novelty of solutions}
\label{sec:novelty}

In this section, we show that the solutions in chapter \ref{sec:aneqb} for the Einstein-Maxwell-dilaton theory with two different coupling constants, which are given by equations \eqref{metric5danoteqb}, \eqref{Atexp} and \eqref{phiexp} are quite novel and can't be uplifted to any known solutions in higher dimensions.

We should note that there are only two known upliftings for the five-dimensional generalized Einstein-Maxwell-dilaton theory with two different coupling constants, In the first uplifting, the solutions can be uplifted to some solutions of the Einstein-Maxwell theory with cosmological constant in higher than five dimensions ($5+{\cal N}$ with ${\cal N}> 0$). In the second uplifting, the solutions can be uplifted to the solutions of six dimensional gravity with a cosmological constant.   

In the first uplifiting process, the coupling constants $a$ and $b$ must be equal  to each other \cite{GK}, \cite{GSSST}, \cite{MYDILPAP}, and moreover the number ${\cal N}$ is equal to 
\be {\cal N}=\frac{3a^2}{1-a^2}.
\ee
However as we notice from equation \eqref{ab}, there is no real value for the coupling constant $a$, such that $a=b$. So, we conclude that the solutions  \eqref{metric5danoteqb}, \eqref{Atexp} and \eqref{phiexp}  to the Einstein-Maxwell-dilaton theory with two different coupling constants, can't be uplifted to any known solutions of the Einstein-Maxwell theory with cosmological constant in higher than five dimensions ($5+{\cal N}$ with ${\cal N} > 0$).

In the second uplifitng process, the coupling constant $a$ must be equal to $\pm 2$ while $b$ must be equal  to  $\pm \frac{1}{2}$, respectively \cite{Steer}, such that they satisfy the equation 
\be 
ab=1,\ee
in complete disagreement with the constraint equation (\ref{ab}). So, we conclude that the solutions  \eqref{metric5danoteqb}, \eqref{Atexp} and \eqref{phiexp}  to the Einstein-Maxwell-dilaton theory with two different coupling constants, can't be uplifted to any known solutions of the six-dimesnional Einstein gravity with cosmological constant. We note 
the metric for the six-dimensional Einstein gravity is given by \cite{Steer}
\begin{equation}
ds_6^2=e^{\mp(\frac{4}{3})(\frac{1}{2})\phi(t,r,\theta)}ds_5^2+e^{\pm4(\frac{1}{2})\phi(t,r,\theta)}(dw+2A_t(t,r,\theta)dt)^2,\label{upl}
\end{equation}
in terms of five-dimensional spacetime $ds_5^2$, the Maxwell field $A_t(t,r,\theta)$  and the dilaton field $\phi(t,r,\theta)$, which are explicitly given in equations  \eqref{metric5danoteqb}, \eqref{Atexp} and \eqref{phiexp}, respectively.
The sixth coordinate $w$ in \eqref{upl} denotes the uplifted coordinate.  We checked explicitly that the metric \eqref{upl} is not a solution to the Einstein's field equations in the presence of a cosmological constant in six dimensions. We also should note that the second uplifting process, that has been proposed in \cite{Steer}, works only for five-dimensional diagonal line element $ds_5^2$ in \eqref{upl} \cite{MYDILPAP}. Hence we should expect that, the uplifting to six dimensions is not possible, since the five-dimensional spacetime \eqref{dsanoteqb} has one off-diagonal term from Nutku line element  \eqref{Nutku}.
 
In conclusion, we can't uplift the solutions 
\eqref{metric5danoteqb}, \eqref{Atexp} and \eqref{phiexp}  to the Einstein-Maxwell-dilaton theory with two different coupling constants, to any solutions of the Einstein-Maxwell or Einstein gravity with cosmological constant in higher than five dimensions.  

We consider the $D$-dimensional gravity with a $q+1$-form potential ${{\cal P}}_{[q+1]}$ in the presence of a cosmological constant ${\Lambda}_D$ \cite{GK} as
\begin{equation}
{\cal S}_{D}=\int d^Dx\sqrt{-g}\{R-\frac{1}{2(q+2)!}{\cal F}_{[q+2]}^2+2\Lambda_D\}\label{HighD},
\end{equation}
where ${\cal F}_{[q+2]}=d{\cal P}_{[q+1]}$ is the $q+2$-form field strength for the $q+1$-form potential ${\cal P}_{[q+1]}$ and $q+1=D-p,\, p>0$.
Dimensionally reducing the $D$-dimensional gravity theory (described by the action \eqref{HighD}) by 
\begin{equation}
ds^2_{D}=e^{-\delta \hat \phi}ds^2_{p+1}+e^{(\frac{2}{\delta (p-1)}-\delta)\hat \phi}dQ^2_{q},
\end{equation}
to $p+1$ dimensions, where $dQ^2_{q}$ is the metric for the $q$-dimensional compactified curved space and $\hat \phi$ is the dilaton field in $p+1$ dimensions. 
The $q+1$-form potential ${\cal P}_{[q+1]}$ de-compactifies as
\begin{equation}
{\cal P}_{[q+1]}={\cal A}_{[1]}\wedge dQ_{q}.
\end{equation}
We then find the $p+1$-dimensional theory is a generalized Einstein-Maxwell-dilaton theory with two different cosmological constants, as 
\begin{equation}
S_{p+1}=\int d^{p+1}x \{ R-\frac{1}{2}(\nabla \hat \phi)^2-\frac{1}{4}e^{\gamma \hat \phi}\hat{\cal F}_{[2]}^2+2\Lambda_D e^{-\delta \hat \phi}+R_{q} e^{-\frac{2}{\delta(p-1)}\hat \phi}\}
\label{EMD2E},\end{equation}
where 
\be
\hat{\cal F}_{[2]}=d{\cal A}_{[1]},
\ee
and $\delta=\sqrt{\frac{2q}{(p-1)(p+q-1)}}$, $\gamma=(2-p)\delta$.  We also note that in \eqref{EMD2E},  $R_{q}$ is the Ricci scalar of the compactified $q$-simensional space \cite{GK}.
Setting $p=4$, \, $\Lambda_D=0$, \, $R_q=-\Lambda$, together with 
 \begin{eqnarray}
 \gamma&=&-\frac{4}{3}\sqrt{\frac{3}{8}}a, \\
 \delta&=&-\frac{3}{2}\sqrt{\frac{3}{8}}\frac{1}{(p-1)b}, \\
 \hat \phi&=&\sqrt{\frac{8}{3}}\phi, \\
 {\cal A}_{[1]}&=&2A_{t}dt,
 \end {eqnarray} in (\ref{EMD2E}), leads to the five-dimensional generalized Einstein-Maxwell-dilaton theory with the action (\ref{act}). We also note that the constraint (\ref{ab}) between the two different coupling constants $a$ and $b$, is automatically satisfied. 
 
 We consider now the uplifting of the solutions in section \ref{sec:aeqb}, that are given by the metric (\ref{metraeqb}), the Maxwell field (\ref{Afix}) and the dilaton field (\ref{dilfix}), respectively. We check explicitly that embedding the metric (\ref{metraeqb}) and the dilaton (\ref{dilfix}), into a $(5+d)$-dimensional metric
\be
ds_{5+d}^2=e^{\frac{4}{3}\sqrt\frac{{d}}{{d}+3}\phi(t,r,\theta)}ds_5^2+e^{\frac{-4}{{d}}\sqrt\frac{{d}}{{d}+3}\phi(t,r,\theta)}d\vec z\cdot d\vec z,\label{highdEM}
\ee
leads to the Einstein-Maxwell theory with the cosmological constant $\Lambda=\frac{9\tau^2}{2}\frac{({d}+3)({d}+4)}{{d}^2}$, where $d$, the dimension of an Euclidean space $E^{d}$, is related to the dilaton coupling constant $a$, by ${d}=\frac{3a^2}{1-a^2}$. We show the directions of the Euclidean space $E^{d}$ by $\vec z=(z_1,\cdots ,z_{ d})$ in (\ref{highdEM}).


\section{Conclusions}

Inspired with the fact that nearly all well-known solutions in Einstein-Maxwell theory have spherical symmetry, in this article, we considered the role of minimal surfaces and instantons in constructing new solutions to five-dimensional  generalized Einstein-Maxwell-dilaton theory in presence of cosmological constant with two coupling constants. The two coupling constants correspond to the coupling of the dilaton field to the Maxwell field, and to the cosmological constant.  We considered different possibilities for the two coupling constants and constructed the exact analytical solutions for the theory. 

First, we considered the theory with two different coupling constants. We found exact analytical solutions for the metric, the Maxwell field and the dilaton field. Moreover, we found that the cosmological constant depends on one of the coupling constants. We also found that the second coupling constant is not independent of the other coupling constant. We explicitly showed that the solutions can't be uplifted to simpler gravity theories, such as Einstein-Maxwell theory in higher dimensions. However, we showed that a higher-dimensional gravity theory with a form-field can be compactified on an internal curved space, to yield the uplifted version of our solutions.  Up to a conformal transformation,  the spacetime is regular everywhere and we presented some physical properties of the solutions. 
 
Second, we considered the theory with equal coupling constants. We used a different metric ansatz and found exact analytical solutions for the spacetime metric, the Maxwell field and the dilaton field. We discussed the physical behaviours of the solutions. We showed that the solutions can be obtained from compatification of the Einstein-Maxwell theory in higher dimensions for only some special values of the coupling constant. 

Third, we considered the limit in which the coupling constants, are both zero.  We found new class of exact solutions to the Einstein-Maxwell theory, as the  Einstein-Maxwell-dilaton  theory reduces in the limit of zero coupling constant. Though the solutions are asymptotically dS in the limit of zero coupling constant, the solutions with non-zero coupling constants are not necessarily dS or anti dS. We leave studying the thermodynamics of these solutions and possible constructing other solutions based on other types of minimal surfaces, for a forthcoming article.

\vskip 2cm

{\Large Acknowledgments}

This work was supported by the Natural Sciences and Engineering Research
Council of Canada. 

\vskip 2cm


\end{document}